\newcommand{\diracslash}[1]{#1\llap{/\kern2pt}}

\newcommand{\be}{\begin{equation}}
\newcommand{\ee}{\end{equation}}
\newcommand{\bea}{\begin{eqnarray}}\index{\footnote{}}
\newcommand{\eea}{\end{eqnarray}}
\newcommand{\ba}[1]{\begin{array}{#1}}
\newcommand{\ea}{\end{array}}

\newcommand{\threej}[6]{
\ensuremath{
\left( \!\!
\begin{array}{ccc}
#1 & #2 & #3 \\%[-20\arrayrulewidth]
#4 & #5 & #6 \\%[-20\arrayrulewidth]
\end{array}
\!\!\right) 
}}

\documentclass[12pt]{iopart}
\usepackage{graphicx}
\begin{document}
\setlength{\topmargin}{0.2in}

\title[anisotropic interaction in photoassociation]{Resonant enhancement of ultracold photoassociation rate by 
electric field induced anisotropic interaction}
\author{Debashree Chakraborty$^1$, Jisha Hazra$^1$ and Bimalendu Deb$^{1,2}$}
\address{$^1$Department of Materials Science, Indian Association for the Cultivation of Science,
 Jadavpur, Kolkata 700032, INDIA}
\address{$^2$Raman Center for Atomic, Molecular and Optical Sciences, Indian Association for the Cultivation of Science,
Jadavpur, Kolkata 700032, INDIA}

\begin{abstract}
We study the effects of a static electric field on the photoassociation of a heteronuclear atom-pair  into a polar molecule.
The interaction of permanent dipole moment with a static electric field largely affects the ground state continuum wave function of the
atom-pair at short separations where photoassociation transitions occur according to Franck-Condon principle. Electric field induced anisotropic interaction
between two heteronuclear ground state atoms leads to scattering resonances at some specific electric fields. Near such resonances the amplitude of scattering wave function
at short separation increases by several orders of magnitude. As a result, photoaasociation rate is enhanced by several orders of magnitude near the resonances. We discuss in detail electric field modified atom-atom scattering properties and resonances. We calculate photoassociation rate that shows giant enhancement
due to electric field tunable anisotropic resonances.
We present selected results among which particularly important are the excitations of higher rotational levels in ultracold photoassociation due to electric field tunable resonances.     
\end{abstract}


\maketitle

\section{Introduction}
 Research interest in the field of cold molecules has witnessed a tremendous growth in recent times. 
The possibility to investigate molecular behaviour at low temperatures motivates physicists and chemists 
from diverse backgrounds to study cold and dense molecular gases. Certainly, this development is inspired
 by the great success in the closely related field of cold atoms. Molecules can have properties which are not 
available with atoms, for instance a hetero-nuclear molecule can possess permanent electric dipole moment. While the rich 
internal structure of molecules possesses new challenges for cooling and trapping them, the potential
applications of cold molecules are remarkable. This is especially true for ultracold polar molecules which 
 allow for a wealth of interesting studies. The interaction of  permanent dipoles with an external 
electric field provides a tool to manipulate physical processes taking place in the ultracold regime. 
Several quantum computation devices have been proposed based on the interaction of the heteronuclear dimers 
with external electric fields \cite{demille,cote}. External electric fields can also alter the internal 
rovibrational structure and dynamics of heteronuclear molecules \cite{gonzalezferez1,gonzalezferez2,gf3}.
The orientation, angular motion, hybridization, 
and changes in the transition rates and lifetimes of the rovibrationally excited state of a LiCs molecule in a 
strong static electric field have been theoretically studied \cite{gf4,gf5}. Electric field is used to enhance the interaction between
Li and Cs atoms in an ultracold collision \cite{krems1,krems2}. This enhancement is due to the interaction of the instantaneous
 dipole moment of the heteronuclear collision complex with the static electric field.
The interaction of
 a heteronuclear pair of atoms with an electric field can couple states of different angular momenta in the electronic ground continuum
of the pair leading to resonances in an ultracold collision.
 
Our purpose here is to utilize the static electric field induced scattering resonances to influence free-bound photoassociation (PA) \cite{rmp} process of molecule formation. 
PA of ultracold atoms via the interaction with an electromagnetic field has become a 
standard technique to produce cold and ultracold molecules. The experimental techniques previously used for 
the formation of homonuclear molecules have been extended to heteronuclear alkali-metal dimers. Several experimental
groups have reported  the photoassociative formation of ultracold alkali-metal dimers , such as NaCs \cite{bigelow},
KRb \cite{marcassa,stwalley,wang}, RbCs \cite{kerman,sage} and LiCS \cite{kraft} in their electronic ground states.
The effect of a static electric field on the formation of heteronuclear molecules in their electronic ground state via one-photon 
stimulated emmision from ground continuum has been theoretically studied \cite{gf6}.
%===========================================================================================================
% Figure- 1
\begin{figure}
\includegraphics[width=4.25in]{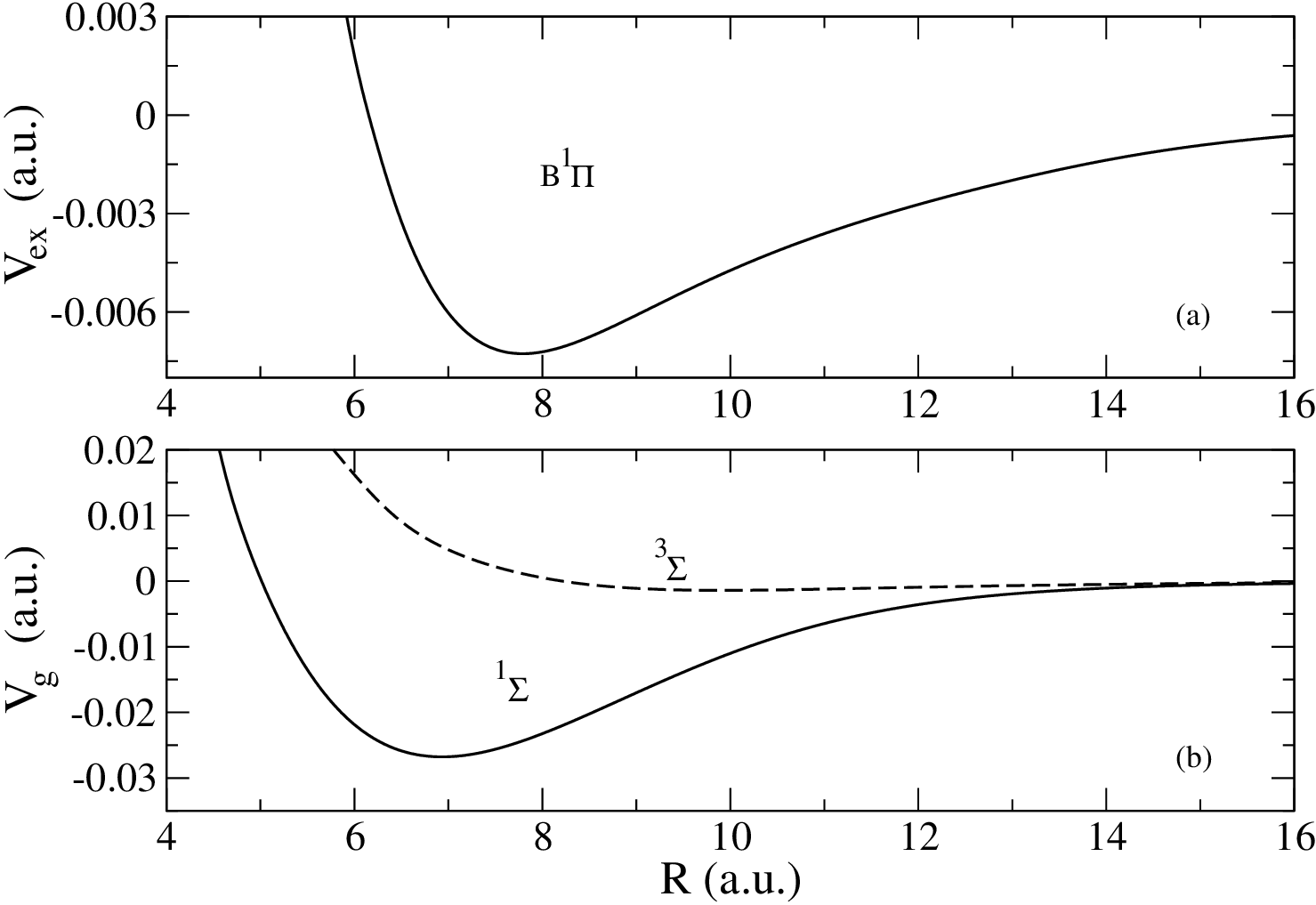}
\caption{Adiabatic Born-Oppenheimer potentials of LiCs without hyperfine interaction for excited (a) and ground (b) states.
The ground state asymptotically corresponds to both Li and Cs atoms in electronic S state while excited state corresponds to
Li in S state and Cs in P state.}
\end{figure}
%============================================================================================================

Here we investigate the effect of a static electric field on the photoassociation process of a 
heteronuclear atom pair to produce molecules in an excited electronic state. Since an electric field can couple
different angular momentum states (partial waves), we need to investigate anisotropic scattering at low energy. 
Taking LiCs molecule as a prototype, we first present a detailed investigation of low-energy  anisotropic atomic collision in the presence of a static electric field.
We have used Numerov-Cooley algorithm-based multichannel scattering techniques and found several anisotropic resonant structures in the 
scattering cross-sections of $^{7}$Li + $^{133}$Cs collision.  
Although the effects of a magnetic Feshbach resonance \cite{rmp1} on PA has been recently studied both experimentally \cite{Junker,Vuletic}
and theoretically \cite{JPBs,JPBs1,MattMack,Cote1,NJP}, to the best of our knowledge, the effects of electric field induced anisotropic resonances 
on PA into excited molecular levels have not been yet studied. One notable feature of the influence of these anisotropic resonances on PA is
the occurrence of higher rotational excitations in excited molecule. This follows from the large modification of two-atom continuum wave functions
for higher partial waves. Owing to the strong spatial dependence of permanent dipole moment at short separations, the interaction of a static electric
field with the permanent dipole moment of a heteronuclear atom-pair leads to large enhancement of continuum wave functions at short separations,
near electric fields at which anisotropic resonances occur.

This paper is organised as follows: First, we briefly discuss how an external static electric field modifies the 
effective interaction potential between two heteronuclear collision pair. In section 2, we present mathemetical
formulation of the problem with an emphasis on anisotropic scattering. The effects of an external static electric field on the 
ground state scattering of $^{7}$Li + $^{133}$Cs are discussed in section 3. In section 4, we discuss the effects of static electric field-induced
anisotropic interaction on PA rate and present our main results. 
Finally we summarize and come to the conclusion in section 5.

\section{Formulation}
%===========================================================================================================
% Figure-2
\begin{figure}
\includegraphics[width=4.25in]{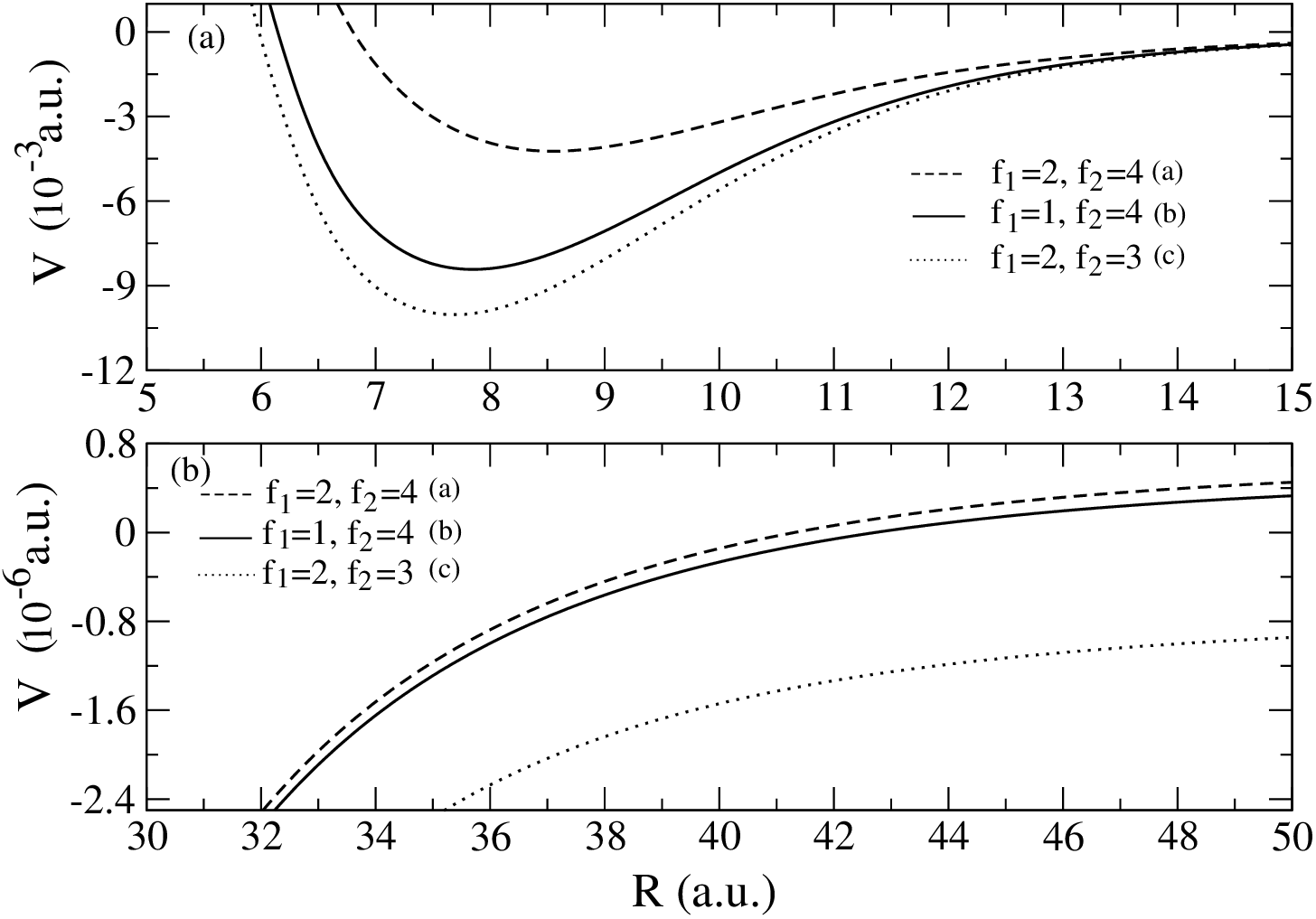}
\caption{The potentials V(R) of LiCs in different diabatic hyperfine channels as a function of internuclear distance R
are plotted. Panel (a) shows the short range and panel (b) shows   
the long range part of the potentials. $f_1$ and $f_2$ are the hyperfine quantum numbers
of $^{7}$Li and $^{133}$Cs, respectively (See Tabel-I in the text).}
\end{figure}
%============================================================================================================
The dynamics of Li-Cs collision in the presence of an electric field is effectively described by the radial 
Hamiltonian
\bea
 \hat H_{R} = -\frac{1}{2\mu R}\frac{\partial^{2}}{\partial R^{2}}R + \frac{\hat L^2(\theta,\phi)}{2\mu R^2} + 
\hat V_c(R) + \hat V_{hf} + \hat V_{\cal E} \eea
where  $\mu = {m_1 m_2}/{(m_1 + m_2)}$
is the reduced mass of two atoms $^{7}$Li and $^{133}$Cs with masses $m_1$ and $m_2$ respectively, R is the interatomic distance, $\hat L^2$ is the 
rotational angular momentum of the collision complex and the angles $\theta$ and $\phi$ specify the orientation 
of the interatomic axis in the space fixed coordinate frame. The electronic interaction potential can be 
represented as 
\bea
\hat V_c(R) = \sum_S\sum_{M_S}\mid S M_S\rangle V_S(R)\langle S M_S\mid\eea 
where S is the total electronic spin of the two atoms and $M_S$ is the projection of S on the Z-axis.
 $V_S(R)$ represents the adiabatic interaction potential of the molecule in the spin state S. The ground state $V_g(R)$ and the excited state $V_{ex}(R)$ potentials as shown in figure 1,
are taken from \cite{stannum} and \cite{wester}, respectively. This interaction is therefore diagonal in the adiabatic basis $\mid IM_I;SM_S \rangle$, 
\bea
\langle S'M_S' ; I'M_I'\mid \hat V_c(R)\mid SM_S ; IM_I \rangle = 
\delta_{I,I'}\delta_{M_I,M_I'}\delta_{S,S'}\delta_{M_S,M_S'}V_S(R)\eea
where $\vec{S} = \vec{s_1} + \vec{s_2}$ and $\vec{I} = \vec{i_1} + \vec{i_2}$, $\vec{s_1}$ 
and $\vec{s_2}$ are the electronic spins while $\vec{i_1}$ and $\vec{i_2}$ are the nuclear spins of the Li 
and Cs atoms, respectively. 
The total hyperfine Hamiltonian $\hat V_{hf}$ for the collision complex can be written
as a sum of the two atomic hyperfine Hamiltonian
\bea
\hat V_{hf} = \sum_{j=1}^{2} \frac{a_{hf}^{(j)}}{\hbar^2} \vec s_j. \vec i_j =
 \sum_{j=1}^{2} \frac{a_{hf}^{(j)}}{2\hbar^2}(\vec f_{j}^{2} - \vec s_{j}^{2} - \vec i_{j}^{2})  \eea
where $\vec f_{j} = \vec s_{j} + \vec i_{j}$ is the total spin and $a_{hf}^{(j)}$ is the hyperfine 
constant of atom $j$ which is 402.00 MHz for $^7$Li and 2298.25 MHz for $^{133}$Cs \cite{metcalf}. This hyperfine interaction is 
diagonal in the atomic or diabatic or long range basis $\mid \ell,f,f_1,f_2 \rangle$ in which $\hat V_c(R)$ can be expressed as, 
\bea
\langle (f_1' f_2')f' m_{f'} \ell' m_{\ell'} \mid \hat V_c(R) \mid (f_1 f_2)f m_{f} \ell m_{\ell} \rangle 
\nonumber\\=  
\sum_{S'M_{S'},I'M_{I'}} V_S\langle S M_S; I M_I; \ell m_{\ell} \mid (f_1 f_2)f m_{f} \ell m_{\ell}   
\rangle\nonumber \\ \times
\langle (f_1' f_2') f' m_{f'}; \ell' m_{\ell'} \mid S' M_{S'}; I' M_{I'}; \ell' m_{\ell'} \rangle. 
\eea
The transformation from coupled hyperfine representation to short range representation is given by
\bea
\langle S M_S; I M_I; \ell' m_{\ell'} \mid (f_1 f_2) f m_f; \ell m_{\ell} \rangle \nonumber\\= 
\delta_{\ell \ell'}  \delta_{m_{\ell} m_{\ell'}} \langle S M_S; I M_I \mid f  m_f \rangle  
\sqrt{(2 f_1 + 1)(2 f_2 + 1)(2 S + 1)(2 I + 1)} \nonumber\\ \times
\left\{ \begin{array}{ccc} s_1 & i_1 & f_1\\ s_2 & i_2 & f_2 \\ S & I & f \end{array} \right\}
\eea 
%=============================================================================================================
% Figure-3
\begin{figure}
 \includegraphics[width=4.25in]{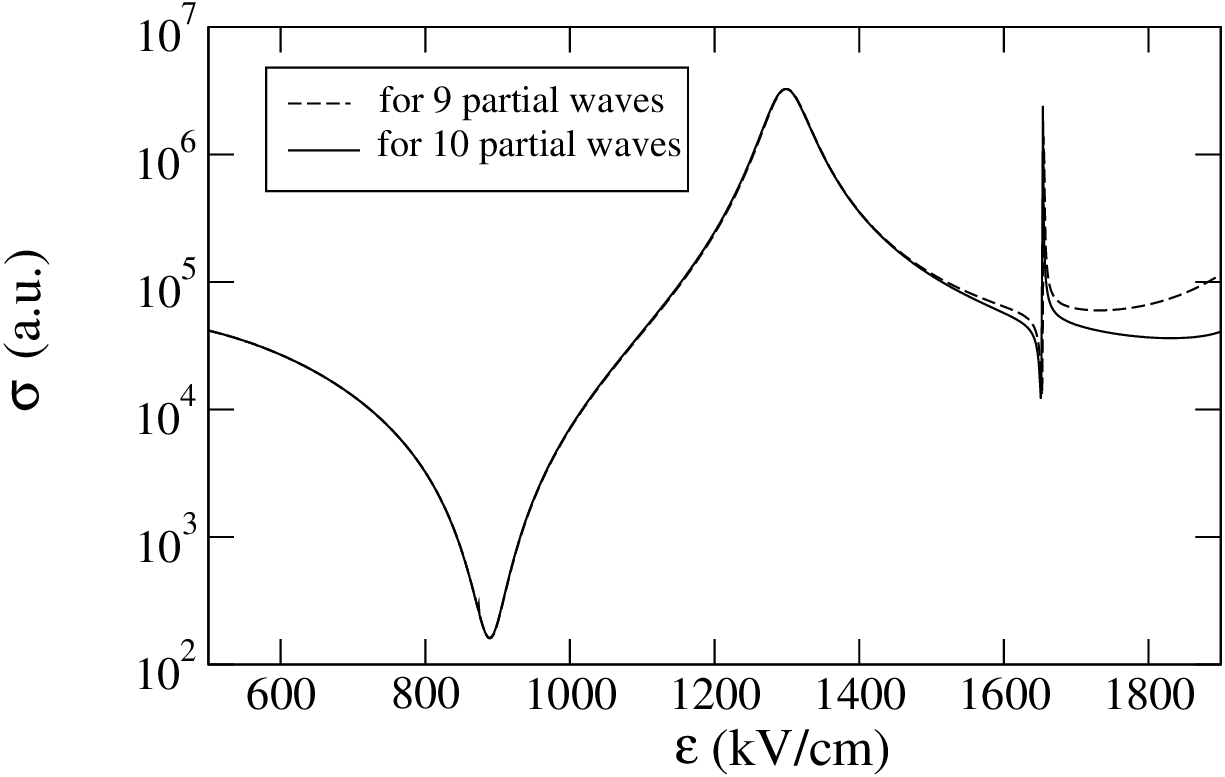}
\caption{Total elastic cross-section $\sigma$ in a.u. vs electric field $\cal E $ in kV/cm is plotted 
for different total number of partial waves to show the convergence of the first resonance point 
at $\cal E$ =  1298 kV/cm.}
\end{figure}
%=============================================================================================================== 
\Table{\label{abrefs}Three different hyperfine channels of $^{7}$Li and $^{133}$Cs.} 
\br
Channel index & $f_1$($^{7}$Li) & $f_2$($^{133}$Cs) & $f$\\
\mr
a & 2 & 4 & 5\\
b & 1 & 4 & 5\\
c & 2 & 3 & 5\\
\br
\endTable
where $\langle S M_S; I M_I \mid f  m_f \rangle$ is Clebsch Gordon coefficient  and the quantity in curly 
braket is 9j-symbol. The hyperfine Hamiltonian can only couples channels with the same total angular 
momentum projection $m_f = m_{f1} + m_{f2} = M_S + M_I$ where $m_{f1}$, $m_{f2}$, $M_S$ and $M_I$ are the
 projections of $f_1$, $f_2$, S and I respectively. For $^7$Li, $i_1 = 3/2$ and for
 $^{133}$Cs, $i_2 = 7/2$ (both with $s_1$ = $s_2$ = 1/2). Since $m_f$ takes values from -6 to +6, the total degeneracy of the atom pair is 128. However, atoms mainly collide on the 
Li($2  {^2{S}_{{1}/{2}}},f_1 = 2$) + Cs($6 {^2{S}_{{1}/{2}}},f_2 = 3$) channel, and thus only 35 degenerate 
entrance channels have to be considered. For $f = 5$, three collisional channels are possible as given in Table 1. The potentials of these
three channels are shown in figure 2 . We consider a single asymptotic 
hyperfine channel (c) with $f_1 = 2, f_2 = 3$  
since the difference in energy of this channel from the other two channels is large enough compared to collision energy at ultracold temperatures. Thus we can approximate our calculation as a single channel 
with $f_1 = 2, f_2 = 3$. The operator $\hat V_{\cal E}(R)$ describes the interaction of the atoms with an external electric 
field. It can be written in the form
\bea
\hat V_{\cal E}(R) = -\cal E \rm cos\theta\sum_S\sum_{M_S}\mid S M_S\rangle d_S(R)\langle S M_S\mid\eea
where d$_S$(R) denotes the dipole moment functions of LiCs in the different spin states and $\cal E$ the electric
 field magnitude. Li and Krems \cite{krems2} have given an analytical expression for this dipole
 moment function approximating the numerical data computed by Aymar and Dulieu \cite{aymer}.  
This analytical expression is given by 
\bea
d_S(R) = D \exp \left[ -\alpha(R - R_e)^2\right], \eea
with the parameters $R_e$ = 7.7 $a_0$, $\alpha$ = 0.1 $a_0^{-2}$, and $D = 6$ Debye for the singlet state
 and $ D = 0.5$ Debye for the triplet state (where $a_0$ = Bohr radius ). The matrix element of $\hat V_{\cal E}(R)$ are 
evaluated using the expressions

\bea
\langle \ell m_\ell\mid\cos\theta\mid \ell' m_{\ell'} \rangle  &=&
 \delta{m_\ell m_{\ell'}} (-1)^{m_\ell}\threej{\ell}{1}{\ell'}{-m_\ell}{0}{m_{\ell'}}\threej{\ell}{1}{\ell'}{0}{0}{0}\\ \nonumber&\times& 
\left[(2\ell+1)(2\ell'+1)\right]^{{1}/{2}} \eea and
\bea
\langle S M_s\mid\left(\sum_{S''}\sum_{M_{S}''}\mid S'' M_{S}''\rangle d_{S''}\langle S'' M_{S}''\mid\right) 
\mid S' M_{S}'\rangle = d_S \delta_{SS'}\delta_{M_SM_{S'}}\eea
where $\langle\vec R\mid\ell, m_\ell\rangle = Y_{\ell m_{\ell}}(\hat R)$. Due to the symmetry of the z-component of the angular momentum, the 
matrix element $\langle \ell m_\ell\mid\cos\theta\mid \ell' m_{\ell'} \rangle$ exists only if $\ell - \ell' = 
\pm 1$ and $m_{\ell} = m_{\ell'}$. 
Therefore electric field can couple even and odd parity channels to each other but not to themselves.

The scattering wave function can be expressed as
\bea \Psi (\vec R) = \frac{1}{R}\sum_{\ell m_{\ell}}\psi_{\ell,m_{\ell}}(R)Y_{\ell m_{\ell}}^*(\hat k)Y_{\ell m_{\ell}}(\hat R) \eea
which has the asymptotic form  
\bea
R \Psi (\vec R)\sim R\exp(i\vec k.\vec R) + f_{\vec k,\vec k'} \exp(ikR),\eea
where $\vec k$ and $\vec k'$ are the incident and the scattered momentum, respectively. The on-shell elastic scattering is then described by 
$f_{\vec k,\vec k'}$, with the scattered momentum $\vec k' = k\hat R$. Expanding the scattering amplitude 
$f_{\vec k,\vec k'}$ into the complete basis $Y_{\ell m_{\ell}}$, we have
\bea
f_{\vec k,\vec k'} = \frac{4\pi}{k}\sum_{\ell,m_{\ell}} T_{\ell m_{\ell}}(\vec k)Y_{\ell m_{\ell}}(\hat R)\eea
%==============================================================================================================
% Figure-4
\begin{figure}
\includegraphics[width=4.25in]{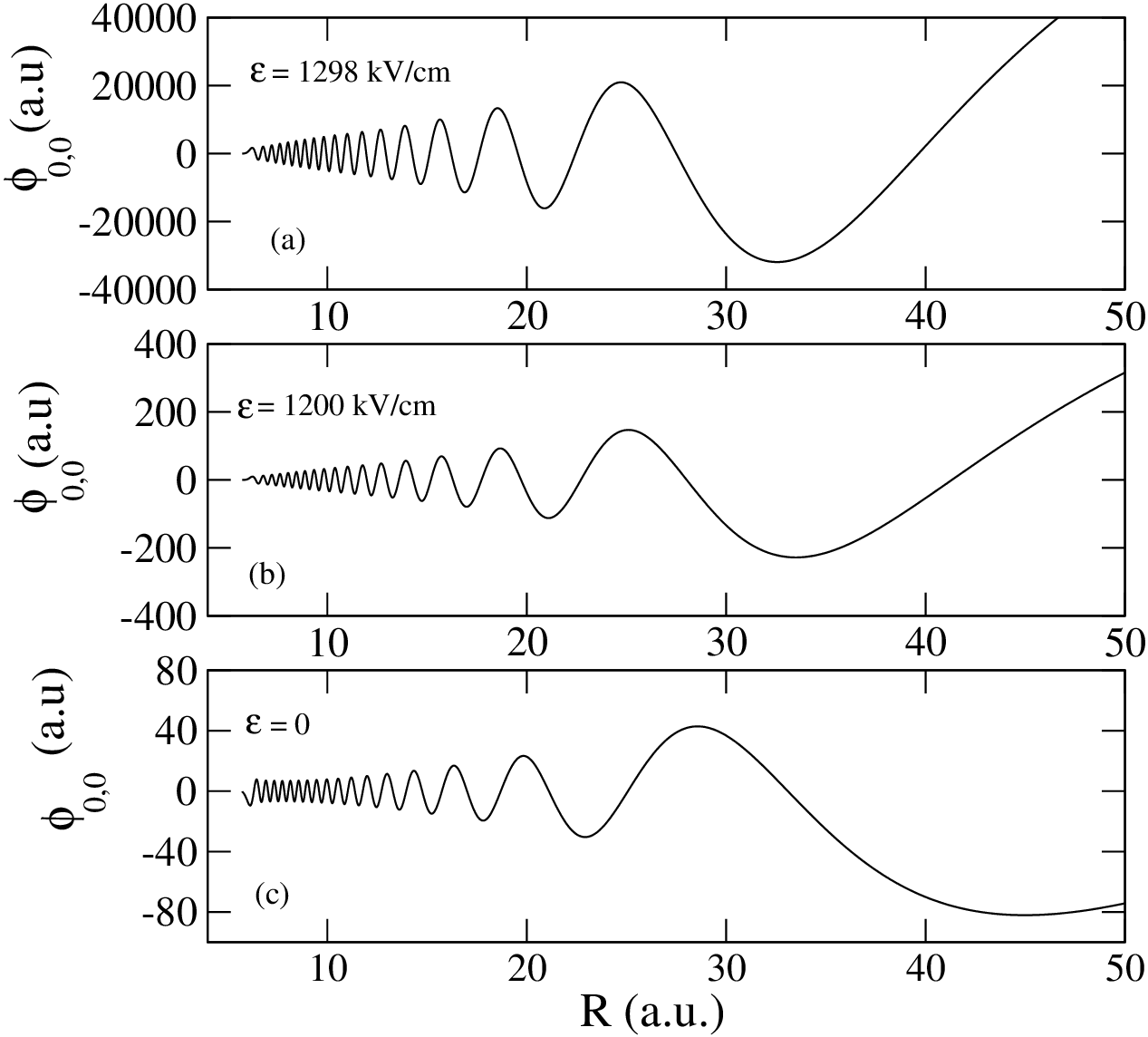}
\caption{Energy-normalised scattering wave function $\phi_{0,0}$ (a.u.) is plotted as 
a function of interatomic distance R (a.u.) at the first resonant electric field $\cal E$ = 1298 kV/cm (a), $\cal E$ = 1200 kV/cm (b) and $\cal E$ = 0 (c). 
}
\end{figure}
%================================================================================================================== 
where $T_{\ell m_{\ell}}$ is a T-matrix element which can also be expanded in the following form
\bea
\frac{1}{k}T_{\ell m_{\ell}}(\vec k) = \sum_{\ell',m_{\ell'}}t_{\ell m_{\ell}}^{\ell' m_{\ell'}}(k)Y_{\ell' m_{\ell'}}(\hat k),\eea
and 
\bea
\exp(i\vec k.\vec R) = 4\pi\sum_{\ell ,m_{\ell}} i^\ell j_\ell(kR) Y_{\ell m_{\ell}}^*(\hat k)Y_{\ell m_{\ell}}(\hat R).\eea
Substituting equations (13) and (15) in equation (12) we get  
\bea
R \Psi(\vec R) &\sim& \frac{4\pi}{k}\sum_{\ell,m_{\ell}} i^\ell [Y_{\ell m_{\ell}}^*(\hat k)\sin(kR - \ell \pi/2)+T_{\ell m_{\ell}}(\vec k)
 \nonumber \\&\times&\exp(ikR - i\ell \pi/2)]  Y_{\ell m_{\ell}}(\hat R)\eea
where the asymptotic form $j_\ell(kr) \sim \sin(kR - \ell \pi/2)/(kR)$ has been used. 
We get the multichannel form of the scattering equation which is given by
\bea
h_\ell \psi_{\ell m_{\ell}}(R) = \sum_{\ell',m_{\ell'}} i^{\ell'-\ell}\langle \ell m_{\ell}\mid V_{\cal E}(R)\mid \ell' m_{\ell'}\rangle
 \psi_{\ell'm_{\ell'}},\eea
where 
\bea
h_\ell = -\frac{\hbar^2}{2\mu}\frac{d^2}{dR^2} + \frac{\hbar^2}{2\mu}\frac{\ell(\ell+1)}{R^2} + V - E.\eea
$E = \hbar^2 k^2 /2\mu$ is the collision energy and V is the central potential for the chosen hyperfine channel. The asymptotic boundary condition on
$\psi_{\ell m_{\ell}}$ can be set as
\bea
\psi_{\ell m_{\ell}} \sim \sin(kR - \ell \pi/2) + T_{\ell m_{\ell}}(k)\exp(ikr - i\ell \pi/2).\eea
%================================================================================================================
% Figure-5
\begin{figure}
 \includegraphics[width=4.25in]{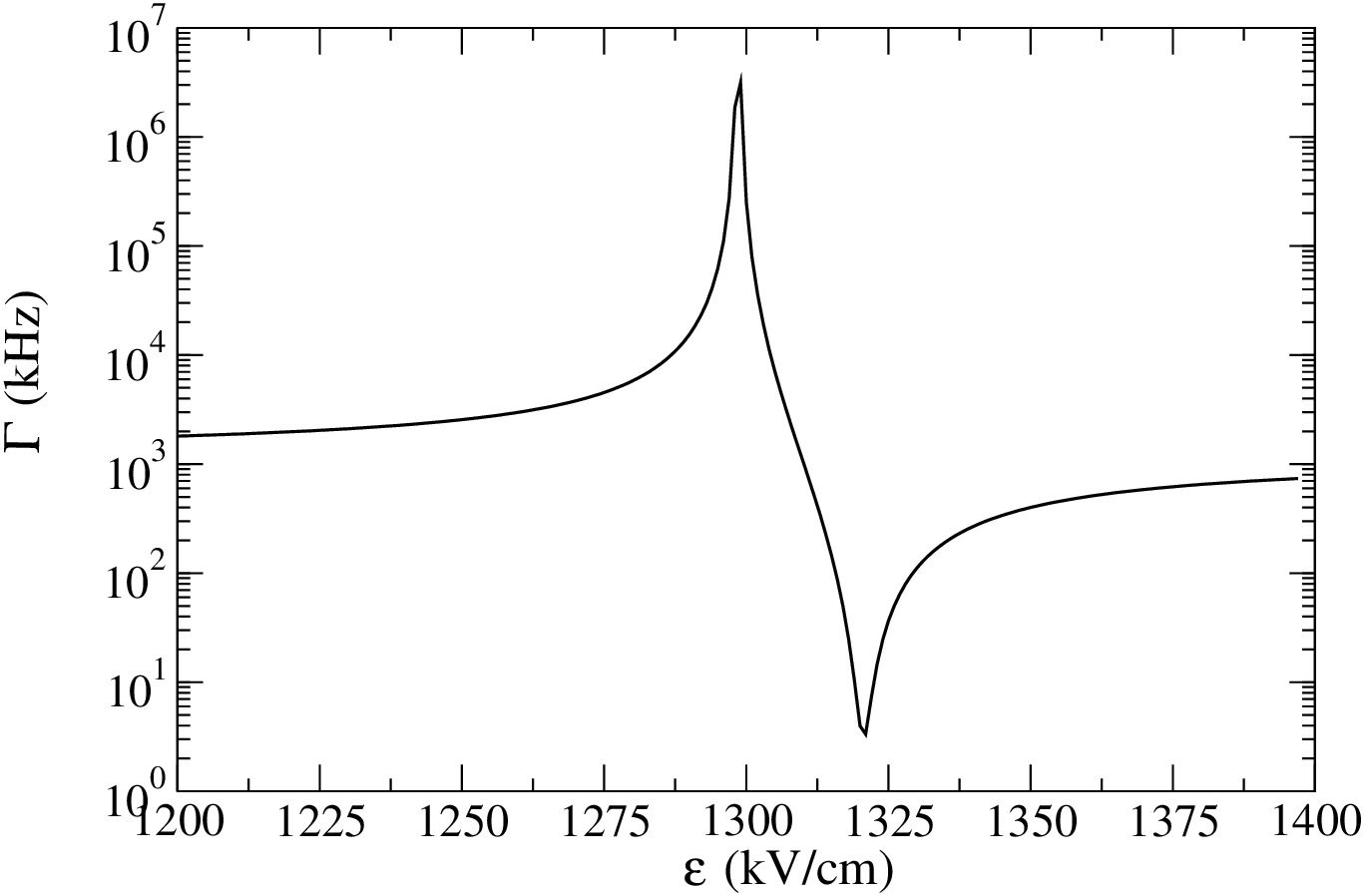}
\caption{Plot of stimulated linewidth $\Gamma$ vs. electric field $\cal E$ at laser intensity 1W/{cm}$^2$.}
\end{figure}
%=================================================================================================================%============================================================
Under the approximation of single asymptotic hyperfine channel as discussed earlier, we can write the coupled equation (17) in a matrix form
in relative angular momentum $(\ell)$ basis as given by 

\bea
\left[\left(-\frac{\hbar^2}{2\mu} \frac{d^2}{dR^2} + {V}\right){\bf I} + \frac{\hbar^2}{2\mu R^2}{\bf L^2 } +
 {{\bf V}_{\cal E}}\right]\bf{\Phi} = {E}{ \bf\Phi }\eea
where ${\bf I}$ is the identity matrix, $\bf L^2 \equiv$ Diag[1(1+1), 2(2+1),......., $\ell$($\ell$+1).....] is a diagonal matrix. The wave function $\bf\Phi$ is a matrix whose elements
are given by $\phi_{\gamma'\gamma}$ where $\gamma\equiv\ell, m_{\ell}$ and $\gamma'\equiv\ell', m_{\ell'}$ are the incident and the scattered angular states respectively.
Therefore the overall wave function for an incident partial wave $\gamma  (\ell,m_{\ell})$ becomes 
\bea
\psi_{\gamma = \ell,m_{\ell}}(E,R) = \sum_{\gamma' = \ell',m_{\ell'},}\phi_{\gamma'\gamma}(E,R)\mid\gamma'\rangle.\eea
By imposing the boundary conditions on the partial waves as given by equation (19), we can get the T-matrix elements, 
$t_{\ell m_\ell}^{\ell' m_{\ell'}}$. The total elastic cross-section is given by
\bea
\sigma = 4\pi\sum_{\ell {\ell'}}\sum_{m_\ell m_{\ell'}}\mid t_{\ell m_\ell}^{\ell' m_{\ell'}}\mid ^2\eea
\section{Ground state scattering: Electric field induced resonances}
To obtain asymptotic scattering solutions, Numerov technique was adopted to numerically propagate equation (20). 
In the asymptotic region, the free wave function goes like
%========================================================================================================
% Figure-6
\begin{figure}
 \includegraphics[width=4.25in]{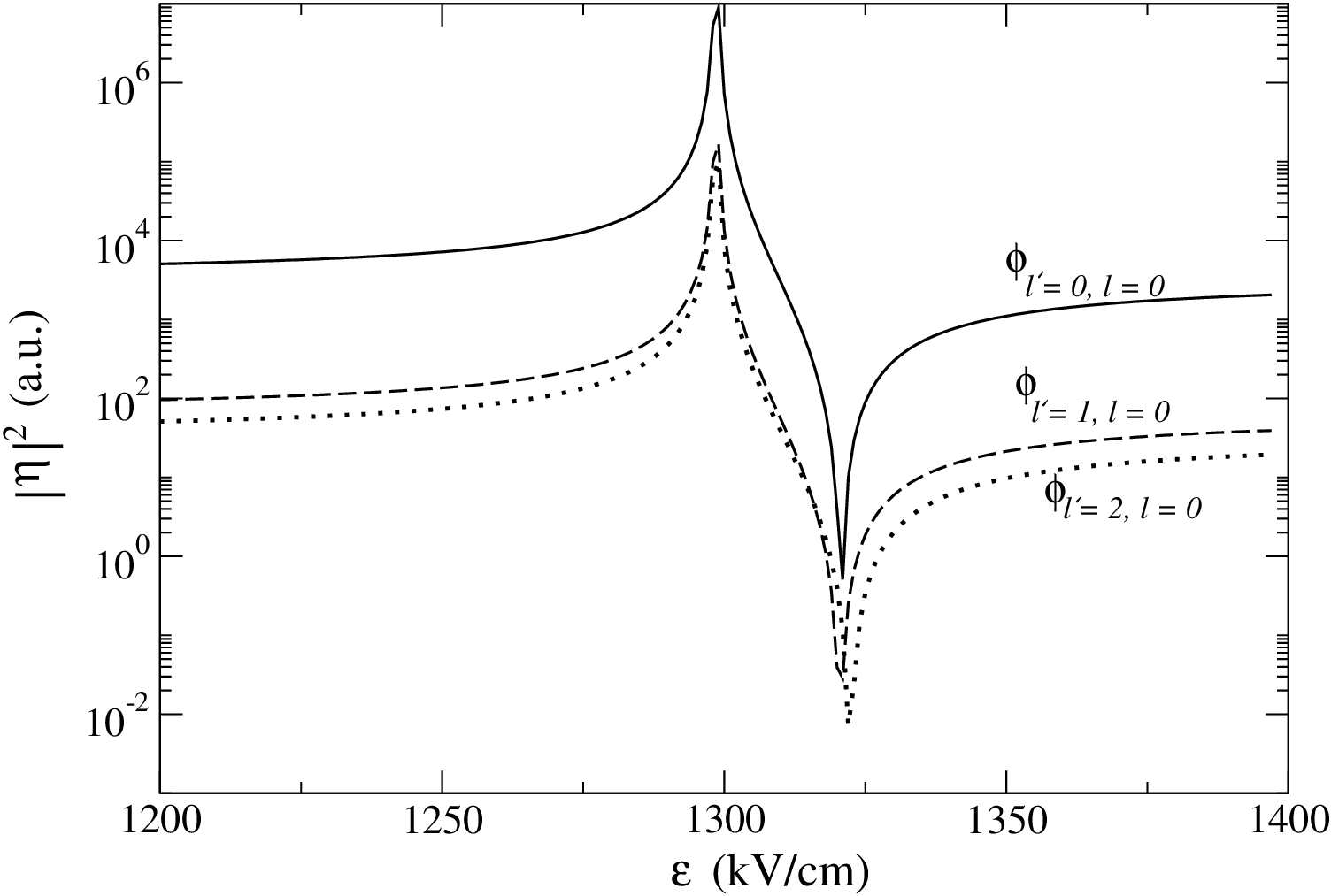}
\caption{Square of franck-condon factor $\mid\eta\mid^{2}$ (a.u.) vs. electric field $\cal E$ (kV/cm) for transitions between the excited 
bound state $v=26$, $J=1$ and different scattering states $\phi_{\ell'0,\ell0}$ with incident partial wave $\ell=0$
and scattered partial wave $\ell'$ = 0 (solid line), $\ell'$ = 1 (dashed line) and $\ell'$ = 2 (dotted line).}
\end{figure}
%=======================================================================================================
\bea
\phi_{l m} = c_1\sin (kR - l\pi/2) + c_2\cos (kR - l\pi/2).\eea
We construct $S$-matrix with its elements given by $(c_1 - ic_2)^{-1} (c_1 + ic_2)$ and the $T$ matrix 
$(S - 1)/(2 i)$. In the absence of electric fields, different partial wave states of the Li-Cs collision complex
are uncoupled and s-wave scattering almost entirely determines the cross-sections at ultralow kinetic energies. The interaction of pemanent dipole 
moment of the colliding pair of heteronuclear atoms with electric field as given by equation (7),
induces coupling between different angular momentum states and may thus affect the ground state scattering wave functions. Ultracold s-wave scattering
is isotropic: the probability to find the atoms after s-wave collisions does not depend on the scattering angle. The interaction with electric fields,
however couples the spherically symmetric s-waves to anisotropic p-wave which in turn is coupled to d-wave, d-wave to g-wave and so on. 
Figure 3 shows anisotropic resonance structures in the scattering cross-section of $^7$Li + $^{133}$Cs
collision at  50$\mu$K energy. Typically convergent results are obtained for a 
minimum angular momentum of $\ell $= 9, as shown in figure 3. Larger the electric field, larger is the number of partial waves required for convergence. We get the first resonance peak near 1298 kV/cm and the second one near 1650 kV/cm.
To show the effect of these
resonances on PA, we have carried out PA calculations near the first resonance peak.

The permanent dipole moment function of the collision complex as given by equation (8) is typically peaked around the equilibrium distance of the diatomic molecule
in the ground state and quickly decreases as the atoms separate. Due to this strong spatial dependence of permanent dipole moment 
at short separations, the interaction of a static electric field with the permanent dipole moment of the collision pair leads to large modification 
of continuum wave functions at short separations. Figure 4 demonstrates that the modification is particularly significant near the resonant electric fields. Since PA transitions take place at short separations 
near the outer turning point of $B^1\Pi$, modification of the continuum wave functions at short separations 
influences PA transition probability as described in the following section 4.   
\section{Photoassociation: Electric field effects}
%========================================================================================================
% Figure-7
\begin{figure}
\includegraphics[width=5.0in]{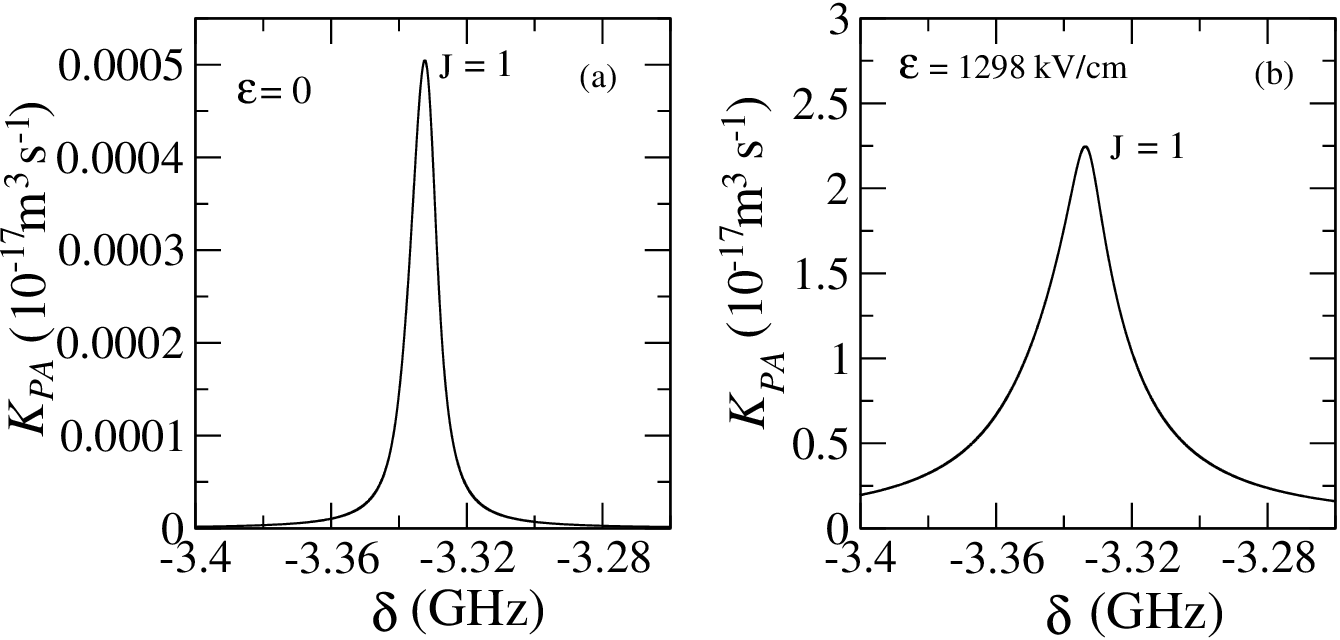}
\caption{The photoassociation rate ${K_{PA}}$ (in unit of 10$^{-17}$cm$^3$s$^{-1}$) is plotted as a function of atom-field 
detuning $\delta$ in GHz $\cal E$ = 0 (a) and at resonant electric field $\cal E$ = 1298 kV/cm.}
\end{figure}
%=======================================================================================================                
Recently, an experiment has demonstrated the formation of ultracold bosonic ${^7}$Li$^{133}$Cs molecule in their 
rovibrational ground state by a two-step 
PA procedure \cite{deiglmayr2}. For illustration of electric field effects on PA, we consider PA transition from Li($2  {^2{S}_{{1}/{2}}},f_1 = 2$) + Cs($6 {^2{S}_{{1}/{2}}},f_2 = 3$) continuum
to the $ v = 26$ , $J = 1$ level of $B^1\Pi$, near the electric fields at which anisotropic resonances occur. Photoassociation rate coefficient \cite{Napolitano} is given by
\bea K_{PA}(T,\omega_L) = \left\langle\frac{\pi v_{rel}}{k^2}\sum_{\ell=0}^{\infty}(2\ell +1)\mid S_{PA}(E,\ell,\omega_L)\mid^2\right\rangle\eea
where $v_{rel} =\hbar k/\mu$ is the relative velocity of the two atoms, $S_{PA}(E,\ell,\omega_L)$ is the S-matrix element for the process of loss of atoms due to PA and $\langle....\rangle$ implies an averaging over thermal velocity distribution.
Assuming Maxwell-Boltzman distribution at temperature T we have, 
\bea K_{PA}(T,\omega_L) = \frac{1}{hQ_{T}}\sum_{\ell=0}^{\infty}(2\ell +1)\int_{0}^{\infty}\mid S_{PA}(E,\ell,\omega_L)\mid^2 e^{(-E/K_BT)} dE\eea
where
\bea \mid S_{PA}(E,\ell,\omega_L)\mid^2 = \frac{\Gamma\gamma_s}{[(\delta+\frac{E-E_{v,J}}{\hbar})^2+((\Gamma+\gamma_s)/2)^{2}]}.\eea 
Here $Q_{T} = ({2\pi\mu K_BT}/{h^2})^{3/2}$ is the translational partition function and $\gamma_s$ is the natural linewidth of the photoassociated level.
$E = {\mu v_{rel}^2}/2$ is the relative kinetic energy of the colliding pair of atoms with reduced mass $\mu$, $E_{v,J}$ is the bound state energy and $\delta = \omega_L-\omega_A$ is the
frequency offset between the laser frequency $\omega_L$ and atomic resonance frequency $\omega_A$. 
The stimulated linewidth $\Gamma$ is given by
\bea \Gamma = \frac{\pi I}{\epsilon_{0}c}\mid\langle \phi_{v,J}\mid D_t(R)\mid \psi_{\gamma = l,m_l}(E,R)\rangle\mid^2\eea
 %========================================================================================================
% Figure-8
\begin{figure}
 \includegraphics[width=4.25in]{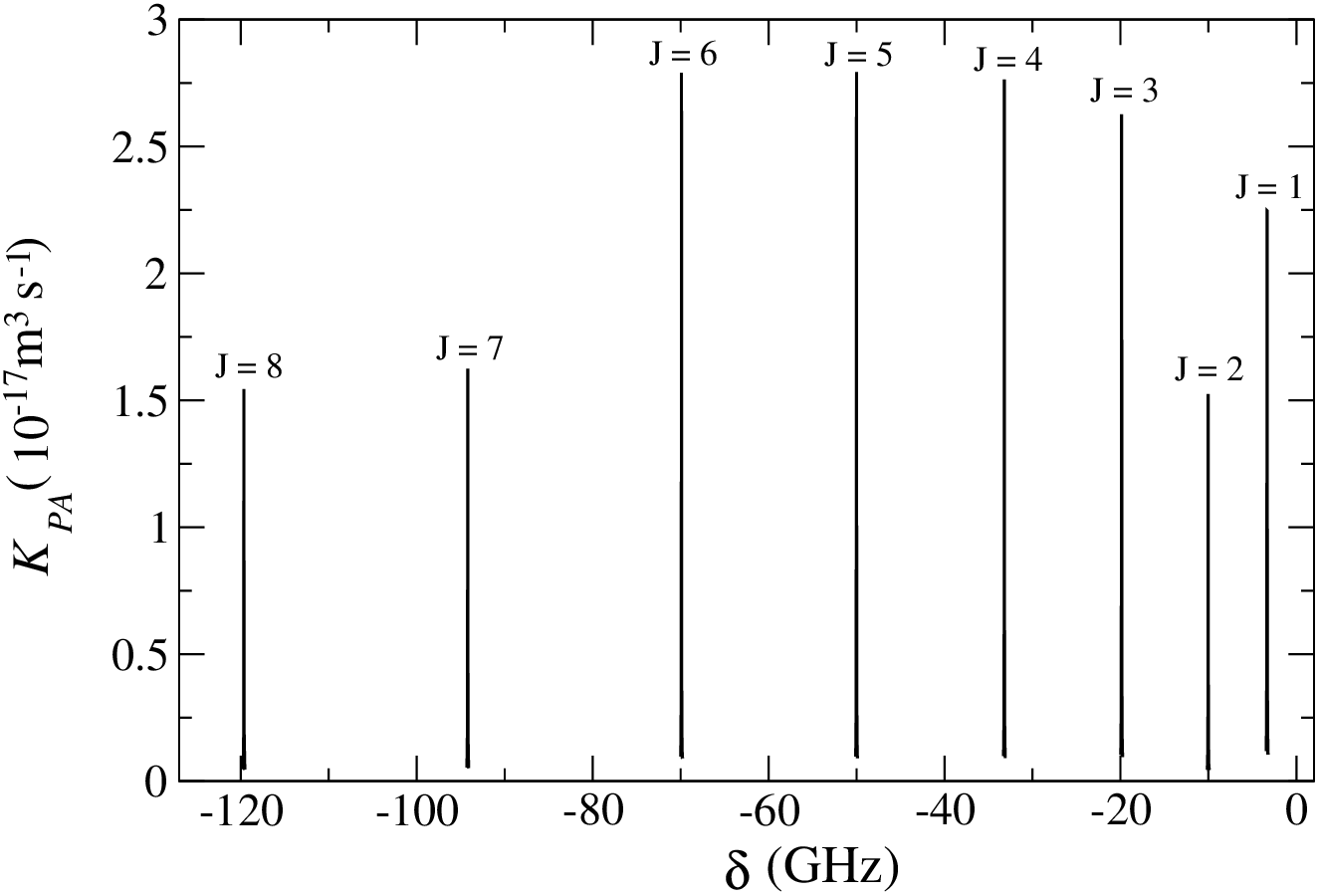}
\caption{The photoassociation rate ${K_{PA}}$ (in unit of 10$^{-17}$cm$^3$s$^{-1}$) is plotted as a function of atom-field 
detuning $\delta$ in GHz at the resonant electric field $\cal E$ = 1298 kV/cm. The rate increases for $J=3,4,5$ and $6$ rotational levels and then start decreasing for $J > 7$.} 
\end{figure}
%=======================================================================================================    
where I is the intensity of the PA laser, $\epsilon_{0}$ and c are the vaccum permitivity and speed of light, respectively. $D_t(R)$ is the transition dipole moment.
At zero electric field  and for 50$\mu$K collisional energy we find $\Gamma$ = 0.315 kHz at laser intensity 1W/{cm}$^2$. Figure 5 demonstrates that the 
stimulated linewidth as a function of electric field $\cal E$ has a resonant structure with giant enhancement near the resonant electric field even at low laser intensity.
This is because of large modifications of continuum wave functions at short separations. To know which components $\phi_{\gamma',\gamma}$ of the continuum wave functions 
have significant contribution to such resonant enhancement of $\Gamma$, we plot square of Franck-Condon (FC) overlap integral $\mid\eta\mid^{2}$ for a few wave component $\phi_{\ell'0,\ell0}$
in figure 6. $\eta$ is defined by
\bea
\eta = \langle \phi_{v,J}\mid \psi_{\gamma = \ell,m_{\ell}}(E,R)\rangle = \langle \phi_{v,J}\mid \sum_{\gamma' =
l',m_l'}\phi_{\gamma'\gamma}(E,R)\mid\gamma'\rangle. \eea
%========================================================================================================
% Figure-9
\begin{figure}
 \includegraphics[width=4.50in]{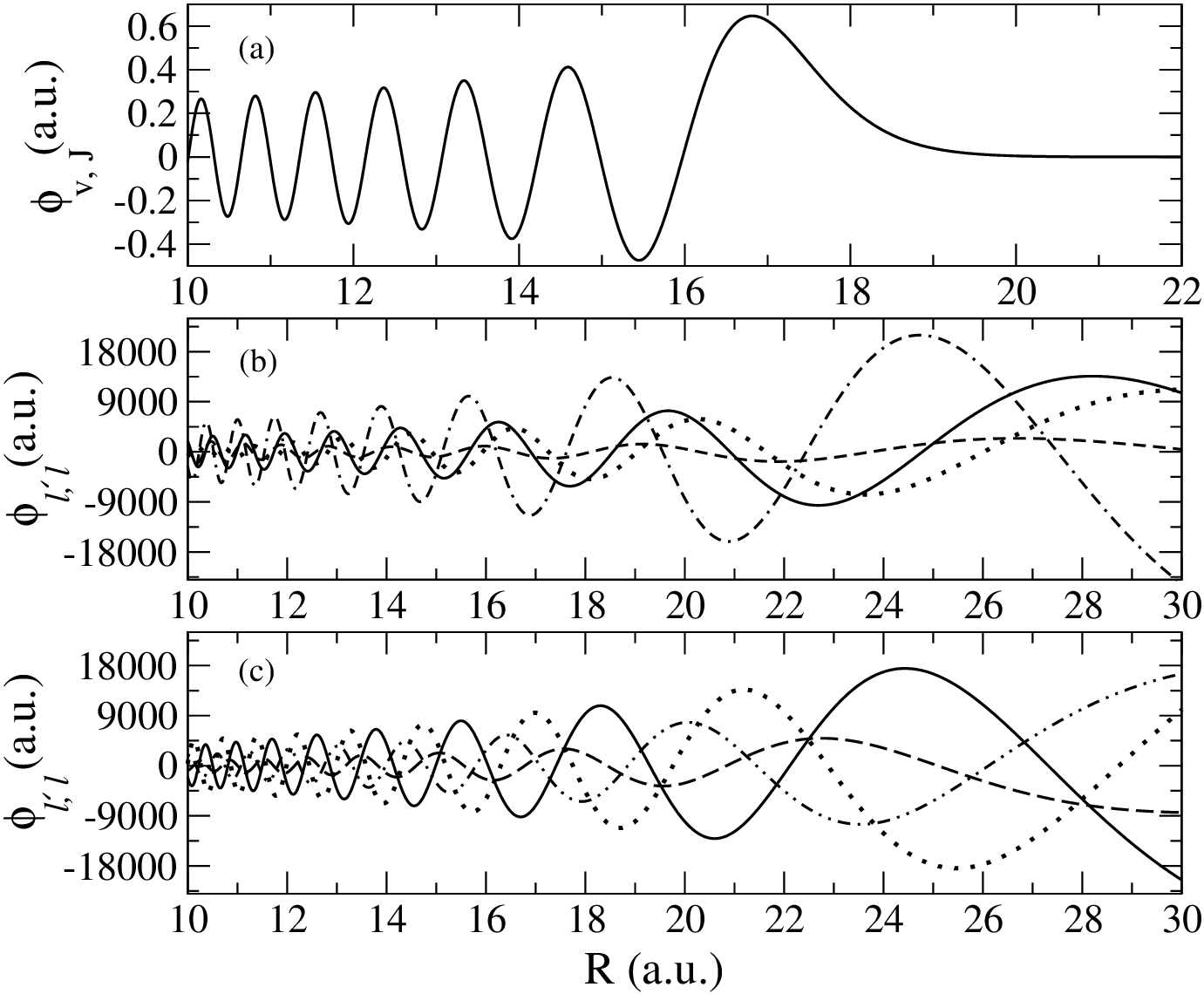}
\caption{(a) The unit normalised bound state $\phi_{v, J}$ in a.u. In panel (b) plotted are the energy normalised scattering wave functions $\phi_{\ell',\ell =0}$ as a function of R (a.u.) for $\ell'$ = 0 (dashed-dotted line), $\ell'$ = 1 (dotted line), $\ell'$ = 2 (solid line) and $\ell'$ = 3 (dashed line).
Panel (c) shows $\phi_{\ell',\ell =0}$ for $\ell'$ = 4 (dotted line), $\ell'$ = 5 (solid line), $\ell'$ = 6 (dashed-dotted line) and $\ell'$ = 7 (dashed line)
at the resonant electric field $\cal E$ = 1298 kV/cm.} 
\end{figure}
%=======================================================================================================
In fact, we find that a large number of partial waves significantly contribute to the resonant features. This is perhaps due to the formation of
a quasibound complex inside the centrifugal barrier of a large number of partial waves which are strongly coupled among them.   
Figure 7 shows that ${K_{PA}}$ near the resonant electric field increases by four orders of magnitude compared to that at zero electric field. This enhancement is due to the electric field induced coupling between different partial waves.
The coupling is so strong that higher rotational levels of the excited electronic state can be populated as shown in figure 8. These higher rotational excitations are not possible at ultracold temperatures in the absence of electric field. The PA rate increases for rotational levels $J=3,4,5$ and $6$ and then starts decreasing for $J>7$ as shown in figure 8.
This can be understood from figure 9 which shows that the amplitude of $\phi_{\ell'=4,\ell=0}$, $\phi_{\ell'=5,\ell=0}$ and $\phi_{\ell'=6,\ell=0}$,wave
functions near the outer turning point of the excited molecular level are larger than $\phi_{\ell'=1,\ell=0}$, $\phi_{\ell'=2,\ell=0}$ and $\phi_{\ell'=3,\ell=0}$ waves.
The selection rule for free-bound transition
$\left | J -|\overrightarrow L+ \overrightarrow S| \right| \le \ell' \le \left | J +|\overrightarrow L+ 
\overrightarrow S| \right|$ allows the scattered partial waves $\ell'$ = 0, 1 and 2 to be accessible for PA with J=1.
Similarly, higher rotational levels $J>1$ become populated due to excitations of higher partial waves in the ground continuum through electric field induced anisotropic scattering.

The enhancement of PA rate near resonant electric fields critically depends on the outer turning point of the molecular bound state that is accessible to the PA transition. As mentioned earlier,
the anisotropic resonances bring about large modification of the amplitude of continuum wave functions at relatively short separations. Now, if a 
prominent antinode of the modified continuum wave functions lies at a separation near the outer turning point of the bound state, then enhancement of PA
rate is expected due to large FC overlap. We have chosen a particular bound state for which enhancement can be achieved. Although, in this paper 
we have studied enhancement of the rate of formation of excited molecular states only, the similar method of electric field induced resonances can readily
be extended to enhance the formation rate of ground molecular states via one photon stimulated emission \cite{gf6}. Enhancement can also be achieved using
quantum interference \cite{JPBs,JPBs1,MattMack1} via laser-coupling another bound state with the continuum. In case of PA in the presence of a magnetic Feshbach 
resonance, a quasi-bound state embedded in the ground continuum gives rise to Fano type
quantum interference \cite{JPBs1} leading to enhancement in PA \cite{Junker}. In case of heteronuclear atoms, quantum interference can be used to
enhance the production of ground molecules due to the existence of permanent dipole moment. Recently, an all optical method of quantum interference
scheme has been proposed for efficient production of ground state polar molecules \cite{MattMack1}. 
The unique feature of the electric field induced enhancement would be the controllability of the excitations of higher rotational states.

\section{Conclusion:}
In conclusion, we have analysed the effects of a static electric field on the PA of a heteronuclear atom-pair. Our results show that it is possible 
to enhance PA rate by several orders of magnitude by tuning electric field near anisotropic resonances. Due to anisotropic nature of the interaction,
a large number of partial waves in the ground continuum become strongly coupled. As a consequence, higher rotational levels can be populated
in an excited dimer formed by PA at ultralow tempertures. This leads to the possibility of selectively populating higher rotational levels in ground
state polar molecule by Raman-type two-color coherent PA. Furthermore it may be interesting to investigate into the effects of electric field on
stimulated Raman adiabatic passage (STIRAP) from free atoms to ground state molecules. Electric field induced resonances may be coupled with
multiple transition pathways to devise novel opto-electrical quantum interference schemes for efficient production of selective rovibrational molecular states.
\section{Acknowledgment}
One of us (Debashree Chakraborty) is grateful to CSIR, Government of India, for a support.

\end{document}